\RequirePackage{ifpdf}
\ifpdf 
\documentclass[pdftex]{sigma}
\else
\documentclass{sigma}
\fi

\begin{document}

\allowdisplaybreaks

\renewcommand{\thefootnote}{$\star$}

\renewcommand{\PaperNumber}{038}

\FirstPageHeading

\ShortArticleName{Integrals Generated by Powers of an Operator}

\ArticleName{First Integrals of Extended Hamiltonians in $\boldsymbol{n+1}$\\ Dimensions Generated by Powers of an Operator\footnote{This paper is a
contribution to the Special Issue ``Symmetry, Separation, Super-integrability and Special Functions~(S$^4$)''. The
full collection is available at
\href{http://www.emis.de/journals/SIGMA/S4.html}{http://www.emis.de/journals/SIGMA/S4.html}}}

\Author{Claudia CHANU~$^\dag$, Luca DEGIOVANNI~$^\ddag$ and Giovanni RASTELLI~$^\ddag$}

\AuthorNameForHeading{C.~Chanu, L.~Degiovanni and G.~Rastelli}

\Address{$^\dag$~Dipartimento di Matematica e Applicazioni, Universit\`a di Milano Bicocca,\\
\hphantom{$^\dag$}~Milano, via Cozzi 53, Italia}
\EmailD{\href{mailto:claudia.chanu@unimib.it}{claudia.chanu@unimib.it}}

\Address{$^\ddag$~Formerly at Dipartimento di Matematica, Universit\`a di Torino,\\
\hphantom{$^\ddag$}~Torino, via Carlo Alberto 10, Italia}
\EmailD{\href{mailto:luca.degiovanni@gmail.com}{luca.degiovanni@gmail.com}, \href{mailto:giorast.giorast@alice.it}{giorast.giorast@alice.it}}

\ArticleDates{Received January 31, 2011, in f\/inal form April 03, 2011;  Published online April 11, 2011}

\Abstract{We describe a procedure to construct polynomial in the momenta f\/irst integrals of arbitrarily high degree for natural Hamiltonians $H$ obtained as one-dimensional extensions
of natural (geodesic) $n$-dimensional  Hamiltonians $L$. The Liouville integrability of $L$ implies the (minimal) superintegrability of $H$. We prove that, as a consequence of natural integrability conditions,  it is  necessary  for the construction that the  curvature of the metric tensor associated with $L$ is constant. As examples, the procedure is applied to one-dimensional $L$, including and improving earlier results, and to two and three-dimensional $L$, providing new superintegrable systems.}


\Classification{70H06; 70H33; 53C21}

\medskip

\rightline{\it Dedicated to Willard Miller Jr.}

\section{Introduction}
It is usual to consider  Hamiltonian systems in $\mathbb E^3$ of the form $H=\frac 12 p_r^2+f(r)+\frac 1{r^2}L$,
where $L=\frac 12(p_\theta^2+\frac 1{\sin(\theta)^2}p_\psi^2)+V(\theta,\psi)$, and $(r,\theta,\phi)$ are spherical coordinates.
The examples are innumerable and we just cite the $n$-body systems of points in one dimension \cite{BKM1} whose Hamiltonian  can be written in a similar way
for some $L$ on $\mathbb S^{n-1}$. In this paper we restrict our attention to Hamiltonians of the form $H=\frac 12 p_u^2+f(u)+\alpha(u)L$ in $n+1$ dimensions that are extensions of a~$n$-dimensional Hamiltonian $L(q^i,p_i)$, by calling $u$ the $(n+1)$-th coordinate and $p_u$ its canonical conjugate momentum.
We will call them {\it extended Hamiltonians} for the given~$L$.
We show how to construct a non trivial f\/irst integral of~$H$ in the form $U^m(G)$, $m\in \mathbb N^*$, where $U$ is a dif\/ferential operator built from the Hamiltonian f\/low of $L$, and $G$ is a suitable function.
If $L$ is a natural Hamiltonian and $G$ is independent of the momenta, the f\/irst integral $U^m(G)$ is a $m$-th degree polynomial in the momenta $(p_u,p_i)$. If the Hamiltonian $L$ is integrable, then automatically $H$ has $n+2$ functionally independent f\/irst integrals and is therefore minimally superintegrable. By requiring certain natural integrability conditions, we show that a necessary condition for the construction is the constant curvature equal to $mc$ of the metric tensor of~$L$, where $c$ is a real constant. In few words, {\it for any non null positive integer~$m$,  given a   $n$-dimensional Riemannian or pseudo-Riemannian manifold of constant curvature equal to $mc$, underlying the geodesic part of the Hamiltonian~$L$,  our procedure can construct a natural Hamiltonian~$H$ in $n+1$ dimensions  with an additional first integral $U^m(G)$ which is polynomial of degree~$m$ in the momenta}. If $L$ includes a non constant scalar potential~$V$, compatibility conditions on $V$ are given in order that the procedure may work, otherwise, all the compatible~$V$ can be determined from a f\/irst-order PDE.
This approach is motivated by the results obtained in~\cite{CDR3}, where $L$ was a one-dimensional Hamiltonian. In that paper, we obtained for $L=\frac 12 p_\psi+V(\psi)$ and $\alpha = u^{-2}$ the additional f\/irst integral
\begin{gather*}
U^m(G)=\left[ p_u+\frac 1{m   u}\left( p_\psi\frac {\partial}{\partial \psi}-\frac  {d V}{d\psi}\frac {\partial}{\partial p_\psi}\right) \right]^m \cos (m \psi+\psi_0)
\end{gather*}
with $V(\psi)=\frac k{\sin ^2 (m\psi+\psi_0)},$ and $G(\psi)=\cos (m \psi+\psi_0)$. The resulting Hamiltonian $H$ can be interpreted as generalization of the three-particle Calogero and Wolfes systems and is maximally superintegrable~\cite{CDR1}. It represents also an instance of the more general Tramblay--Turbiner--Winternitz systems \cite{TTW1,TTW2} whose superintegrability has been proved by another way in \cite{KMK3}. In recent years much research has been done about superintegrable natural Hamiltonians with polynomial f\/irst integrals of high degree in the momenta, both classical an quantum. Many results concern systems which admit separable Hamilton--Jacobi equations and the separability of those systems is explicitly employed to build the high-degree f\/irst integrals, see for example~\cite{KMK1,KMK2,KMK3,MPY,Ts}.
The procedure presented here to build the additional f\/irst integral does not assume separability anywhere. However, the requirement of working with constant curvature manifolds allows to consider the extensions of separable Hamiltonian systems, because constant curvature manifolds are the most natural environment for those systems (see for example~\cite{Kl}). It is not by accident that some of the examples of superintegrable Hamiltonians that we provide here in the section of two and three-dimensional systems  reveal themselves to be extensions of separable systems.

\section{Main results}

Let us consider a Poisson manifold $M$ and a one-dimensional manifold $N$.
For any Hamiltonian function $L\in\mathcal{F}(M)$ with Hamiltonian vector f\/ield $X_L$, we consider its extension on $\tilde{M}=T^*N\times M$ given by the Hamiltonian
\begin{gather}\label{HamExt}
H=\frac{1}{2} p_u^2+\alpha(u)L + f(u),
\end{gather}
where $(p_u,u)$ are canonical coordinates on $T^*N$. The Hamiltonian f\/low of (\ref{HamExt}) is
\begin{gather*}
X_H=p_u\frac{\partial}{\partial u}-(\alpha' L+f')\frac{\partial}{\partial p_u}+\alpha X_L,
\end{gather*}
where prime denotes total derivative w.r.t.\ the corresponding variable.

It is immediate to see that any f\/irst integral of $L$ is also a constant of motion of $H$, when considered as a function on $\tilde{M}$.

We want to determine on $L$, $\alpha$ and $f$ necessary and suf\/f\/icient conditions for the existence of two functions $\gamma\in\mathcal{F}(N)$ and $G\in\mathcal{F}(M)$
such that, given the dif\/ferential operator
\begin{gather}\label{U}
U=p_u+ \gamma(u) X_L,
\end{gather}
the function $F$ obtained applying $m$ times $U$ to $G$
\begin{gather*}
F=U^m(G)
\end{gather*}
is an additional f\/irst integral for $H$.
\begin{remark}
We observe that the operator (\ref{U}) is tensorial with respect to transformations acting on $M$ only i.e.\ preserving the form of the extended Hamiltonian (\ref{HamExt}). Moreover $U$ is an injective linear operator on the space of functions polynomial in the momentum $p_u$: indeed if $F\in\mathcal{F}(\tilde{M})$ is polynomial in $p_u$ then $U(F)=0$ if and only if $F=0$ because $U(F)$ is a polynomial of degree higher than $F$.
\end{remark}
\begin{proposition}
For any $L$, $\alpha$, $f$, we have that $X_H U^m(G)=0$ for a function $G(q^i,p_i)$ if and only if
\begin{gather}\label{E1}
(m\gamma'+\alpha)X_L(G)=0,
\\
\label{E2}
\alpha\gamma X_L^2(G)-m(\alpha'L+f')G=0.
\end{gather}
\end{proposition}
\begin{proof}
Being
\begin{gather*}
[X_H,U]=p_u\gamma'X_L-L\alpha'-f',
\end{gather*}
we have $[[X_H,U],U]=0$.
Hence, we can apply the following formula for the power of a~dif\/feren\-tial operator:
\begin{gather*}
X_H U^m = mU^{m-1}[X_H,U]+U^m X_H = U^{m-1}(m[X_H,U]+U X_H),
\end{gather*}
Since we have
\begin{gather*}
UX_H  =  p_u^2\frac{\partial}{\partial u}+p_u\left[\alpha X_L-(\alpha'L+f')\frac{\partial}{\partial p_u}+\gamma X_L\frac{\partial}{\partial u}\right] + \alpha\gamma X_L^2-\gamma (\alpha'L+f')X_L\frac{\partial}{\partial p_u},
\end{gather*}
and, in particular, for any $G(q^i,p_i)$,
\begin{gather*}
UX_H(G)=\alpha\big(p_u X_L(G)+\gamma X_L^2(G)\big),
\end{gather*}
thus, we get for $X_HU^m (G)$ the following expression
\begin{gather*}
U^{m-1}\big(p_u(m\gamma'+\alpha)X_L(G)+\alpha\gamma X_L^2(G)-m(\alpha'L+f')G\big).
\end{gather*}
Since the operator $U$ is injective on the space of functions polynomial in $p_u$ we have the thesis.
\end{proof}

The conditions (\ref{E1}) and (\ref{E2}) are quite dif\/f\/icult to handle in the general case, their analysis can be further pursued by adding some extra hypotheses on the manifold $M$ and on the Hamil\-to\-nian~$L$.

\begin{theorem}\label{Teo0}
Let  $Q$ be a $n$-dimensional $($pseudo-$)$Riemannian manifold with metric tensor~$\mathbf g$.
The natural Hamiltonian $L=\frac{1}{2} g^{ij}p_ip_j+V(q^i)$ on $M=T^*Q$ with canonical coordinates~$(p_i,q^i)$ admits an extension~$H$ in the form \eqref{HamExt}
with a first integral $F=U^m(G)$ with $U$ given by~\eqref{U} and~$G(q^i)$,  if and only if the following conditions hold:
\begin{enumerate}\itemsep=0pt

\item[$1)$]
the functions $G$ and $V$ satisfy
\begin{gather}\label{HessTeo}
 \mathbf{H}(G) + mc  \mathbf{g}G=\mathbf 0, \qquad c\in \mathbb R,
\\
\label{VTeo} \nabla V \cdot \nabla G -2m(cV+L_0)G=0, \qquad L_0 \in \mathbb R,
\end{gather}
where $\mathbf H(G)_{ij}=\nabla_i\nabla_jG$ is the Hessian tensor of $G$.
\item[$2)$] 
for $c=0$ the extended Hamiltonian $H$ is
\begin{gather*}
H=\frac{1}{2}p_u^2+mA(L+V_0)+B(u+u_0)^2,
\end{gather*}
for $c\neq 0$ the extended Hamiltonian $H$ is
\begin{gather*}
H=\frac{1}{2}p_u^2+\frac{m(cL+L_0)}{S^2_\kappa(cu+u_0)}+W_0,
\end{gather*}
with $\kappa, u_0,V_0, W_0, A\in \mathbb R$, $A\neq0$, $B=mL_0A^2$ and
\begin{gather*}
S_\kappa(x)=\left\{\begin{array}{ll}
\dfrac{\sin\sqrt{\kappa}x}{\sqrt{\kappa}}, & \kappa>0, \\
x, & \kappa=0, \\
\dfrac{\sinh\sqrt{|\kappa|}x}{\sqrt{|\kappa|}}, & \kappa<0.
\end{array}\right.
\end{gather*}
\end{enumerate}
\end{theorem}

For the equations (\ref{HessTeo}) it is natural to write integrability conditions. It follows (Lemma~\ref{PropClaudia}) that the maximal dimension of the space of solutions of equation
(\ref{HessTeo}) alone is $n+1$ and is achieved only in the constant curvature
case. We call {\it complete} the solutions $G$ of (\ref{HessTeo}) satisfying these integrability conditions. Therefore, we can restate Theorem \ref{Teo0} as follows.

\begin{theorem}\label{Teo}
Let  $Q$ be a $n$-dimensional $($pseudo-$)$Riemannian manifold with metric tensor~$\mathbf g$.
The natural Hamiltonian $L=\frac{1}{2} g^{ij}p_ip_j+V(q^i)$ on $M=T^*Q$ with canonical coordinates $(p_i,q^i)$ admits an extension~$H$ in the form~\eqref{HamExt}
with a first integral $F=U^m(G)$ with $U$ given by~\eqref{U} and~$G(q^i)$  a complete solution of~\eqref{HessTeo},  if and only if
$Q$ is a $($pseudo$)$-Riemannian manifold with constant curvature $K=mc$ and~\eqref{VTeo} holds.
\end{theorem}
Clearly, in the case of Theorem~\ref{Teo}  the extended Hamiltonian $H$ is again in the form given by item $2$ of Theorem~{\rm \ref{Teo0}}.

Depending on the form of $V$, the space of
common solutions of the two equations~(\ref{HessTeo}) and~(\ref{VTeo}) can be of lower
dimension but, in any case, the maximal number of independent
solutions is obtained in the constant curvature case. Since~(\ref{HessTeo}) and~(\ref{VTeo}) are linear dif\/ferential conditions on~$G$, their common solutions form a linear
space of dimension lower or equal to $n+1$ (dimension of the space solution of (\ref{HessTeo}) alone) parametrized by up to $n+1$ real parameters $(a_1,\ldots,a_{n+1})$; the linear injective operator $U^k$, $k\in \mathbb N^*$, maps this space in a new linear space, with the same dimension, spanned by the functions $U^k(G)$. These functions are f\/irst integrals of $H$ if and only if $k=m$. It is possible that functions~$U^m(G)$ corresponding to dif\/ferent choices of the parameters $(a_j)$ are simultaneously f\/irst integrals of $H$ and functionally independent even for a non-trivial potential~$V$. The analysis of this interesting case and of the relations among f\/irst integrals in general will be done in some future paper. The condition (\ref{VTeo}) can be read as a~condition on the potentials~$V$, whose solutions form a  functional space of potentials depending on (at most) $n+1$ parameters and admitting at least one non zero~$G$ such that $F=U^m(G)$ is a~f\/irst integral. Examples of potentials~$V$ depending on some of the parameters $(a_j)$ are given in the next section. The presence of parameters into the potentials $V$ of superintegrable systems  can allow the application of the St\"ackel transform or, more generally, of the coupling constant metamorphosis to obtain new superintegrable systems, see for example~\cite{KMst} and~\cite{BS}.

\begin{remark}
Equation (\ref{HessTeo}) could admit in particular cases non-vanishing solutions also on manifold with non-constant curvature, provided they depend on less than $n+1$  parameters and, as appears from some example,  do not depend on some of the variables $(q^i)$. A detailed analysis of this case is in progress.
\end{remark}

The proofs of the theorems follow from the next three lemmas.

\begin{lemma} \label{lem1}
If $L$ is a natural Hamiltonian and $G$ a function of $(q^i)$ we have $X_HU^m(G)=0$~-- with $H$ and $U$ defined as in \eqref{HamExt} and
\eqref{U}~--  if and only if the functions $\alpha$, $\gamma$, $f$, $G$ and $V$ satisfy the following differential conditions
\begin{gather}
  \alpha=-m\gamma', \label{E3}\\
 \label{GammaTeo}
\big(\gamma'+c\gamma^2\big)'=0,\\
  \nabla_i\nabla_jG=-mcGg_{ij}, \label{HessLem}
\\
\nabla_i V \nabla^i G =2m(cV+L_0)G, \label{E10}\\
f=mL_0\gamma^2+f_0, \label{Ef}
\end{gather}
where $c$, $L_0$ and $f_0$ are arbitrary constants.
\end{lemma}

\begin{proof}
By using the (pseudo-)Riemannian structure of $Q$ and covariant derivatives we have
\begin{gather*}
X_L = p_i\nabla^i-\nabla_iV\frac{\partial}{\partial p_i}, \\
X_L^2=p_ip_j\nabla^i\nabla^j-\nabla_iV\nabla^i-2p_j\nabla_iV\nabla^j\frac{\partial}{\partial p_i} - p_i\nabla^i\nabla_jV\frac{\partial}{\partial p_j}+\nabla_iV\nabla_jV\frac{\partial^2}{\partial p_i\partial p_j}.
\end{gather*}
Hence, if $G$ does not depend on the momenta, we have
\begin{gather*}
X_L(G)=p_i\nabla^iG,\qquad X_L^2(G)=p_ip_j\nabla^i\nabla^jG-\nabla_iV\nabla^iG.
\end{gather*}
By substituting in (\ref{E1}) and excluding the trivial case of constant $G$ we obtain (\ref {E3}) and the two conditions
\begin{gather}
 \alpha\gamma\nabla_i\nabla_jG=\frac{m}{2}\alpha'g_{ij}G, \label{E4}\\
 \alpha\gamma\nabla_iV\nabla^iG+m(\alpha' V+f')G=0. \label{E6}
\end{gather}
From (\ref{E4}), by separating terms in $u$ from the other ones we get (\ref{HessLem}) and
\begin{gather}
\alpha'=-2c\alpha\gamma, \label{E7bis}
\end{gather}
where $c$ is a real constant.
By substituting (\ref{E3}) in the relation (\ref{E7bis}) we obtain $\gamma''+2c\gamma \gamma'$ that is~(\ref{GammaTeo}).
By~(\ref{E7bis}) the equation~(\ref{E6}) is equivalent to
\begin{gather*}
\alpha\gamma(\nabla V \cdot \nabla G-2mcVG)+mf'G=0,
\end{gather*}
which by separating terms in $u$ from the others and substituting \eqref{E3} splits into (\ref{E10}) and
\begin{gather*}
f'=-2L_0\alpha\gamma=2mL_0\gamma\gamma', 
\end{gather*}
with $L_0$ constant, that integrated gives (\ref{Ef}).
\end{proof}

\begin{remark}
The addition of  a term  $\phi(u)$ to $U$ does not induce weaker constraints on the func\-tion~$f(u)$ in~(\ref{HamExt}):
a straightforward calculation shows that $\phi(u)$ must necessarily vanish.
\end{remark}

Condition (\ref{GammaTeo}) of  Lemma \ref{lem1} determines the possible forms of the extension. The function $\gamma$ and the related extended Hamiltonian are given by the next lemma and are expressed by using the ``tagged'' trigonometric functions
\begin{gather*}
C_\kappa(x)=\left\{\begin{array}{ll}
\cos\sqrt{\kappa}x, & \kappa>0, \\
1, & \kappa=0, \\
\cosh\sqrt{|\kappa|}x, & \kappa<0,
\end{array}\right.
\qquad
S_\kappa(x)=\left\{\begin{array}{ll}
\dfrac{\sin\sqrt{\kappa}x}{\sqrt{\kappa}}, & \kappa>0, \\
x, & \kappa=0, \\
\dfrac{\sinh\sqrt{|\kappa|}x}{\sqrt{|k|}}, & \kappa<0,
\end{array}\right.
\\
T_\kappa(x) = \frac{S_\kappa(x)}{C_\kappa(x)},\qquad
CT_\kappa(x) = \frac{C_\kappa(x)}{S_\kappa(x)}
\end{gather*}
already employed in \cite{Ranada1,Ranada4,MPY}. These function share almost all the properties of the standard trigonometric and hyperbolic functions, conveniently modif\/ied with the tag $\kappa$, as some straightforward calculations will show.

\begin{lemma} \label{Ex}
The equation $\gamma''+2c\gamma\gamma'=0$ for $c=0$  reduces to $\gamma''=0$, its solution is $\gamma=-A(u+u_0)$ with $u_0$ and $A\neq0$ constants. The related extended Hamiltonian is
\[
H=\frac{1}{2}p_u^2+mA(L+V_0)+B(u+u_0)^2,
\]
with $u_0,V_0, A\in \mathbb R$, $A\neq0$, $B=mL_0A^2$, i.e.\ the sum of $L$ and the Hamiltonian of a harmonic oscillator.

The equation $\gamma''+2c\gamma\gamma'=0$ for $c\neq0$  is equivalent to $\gamma'=-c(\gamma^2+\kappa)$ with $\kappa$ a constant, its solution is:
\[
\gamma=\frac{1}{T_\kappa(cu+u_0)}
\]
and the related extended Hamiltonian is:
\[
H=\frac{1}{2}p_u^2+\frac{m(cL+L_0)}{S_\kappa^2(cu+u_0)}+W_0,
\]
where $W_0\in \mathbb R$.
\end{lemma}

\begin{proof}
The f\/irst part of the proposition is trivial: equations \eqref{E3} and \eqref{Ef} give $\alpha=mA$ and $f=mL_0A^2(u+u_0)^2+f_0$. The constant $A$ must be assumed dif\/ferent from zero to avoid the vanishing of $\alpha$. The given form of $H$ is obtained by setting $B=mL_0A^2$ and $V_0=\frac{f_0}{mA}$.

The second part immediately follows from the fact that if $CT_\kappa^{-1}(x)$ is the inverse of the ``tagged'' cotangent $CT_\kappa(x)$ a straightforward computation gives
\[
\frac{d}{dx}CT_\kappa^{-1}(x)=-\frac{1}{x^2+\kappa}\qquad\mbox{hence}\qquad\int\frac{dx}{x^2+\kappa}=-CT_\kappa^{-1}(x).
\]
Recalling that
\[
\frac{d}{dx}CT_\kappa(x)=-\frac{1}{S_\kappa^2(x)}
\]
and $C^2_\kappa(x)+\kappa S^2_\kappa(x)=1$ the equations \eqref{E3} and \eqref{Ef} give
\begin{gather*}
\alpha = \frac{mc}{S_\kappa^2(cu+u_0)},\qquad
f = \frac{mL_0}{S_\kappa^2(cu+u_0)}+f_0-m\kappa L_0.
\end{gather*}
The given form of $H$ is obtained by setting $W_0=f_0-m\kappa L_0$.
\end{proof}

\begin{remark} \label{Ex1}
The functions $\gamma$ and $\alpha=-m\gamma'$ in the non trivial case $c\neq0$ take essentially three dif\/ferent forms that can be summarized by the three values $\kappa=0$, $\kappa=1$ and $\kappa=-1$. These three cases are described in the following table
\begin{center}
\begin{tabular}{|c|c|c|}
\hline
$\kappa=0$ & \tsep{8pt}\bsep{8pt}$\displaystyle{\gamma=\frac{1}{cu+u_0}}$ & $\displaystyle{\alpha=\frac{mc}{(cu+u_0)^2}}$ \\
\hline
$\kappa=1$ & \tsep{8pt}\bsep{9pt} $\displaystyle{\gamma=\frac{1}{\tan(cu+u_0)}}$ & $\displaystyle{\alpha=\frac{mc}{\sin^2(cu+u_0)}}$ \\
\hline
$\kappa=-1$ & \tsep{8pt}\bsep{9pt} $\displaystyle{\gamma=\frac{1}{\tanh(cu+u_0)}}$ & $\displaystyle{\alpha=\frac{mc}{\sinh^2(cu+u_0)}}$ \\
\hline
\end{tabular}
\end{center}
\end{remark}

\begin{lemma}\label{PropClaudia}
For $n\geq 2$ the linear PDE \eqref{HessLem} admits a $(n+1)$-dimensional linear space of solutions, if and only if the metric $\mathbf g$ has constant curvature $K=mc$.
In particular the trivial case $c=0$ is possible only on a locally flat manifold.
\end{lemma}

\begin{proof}
In components the equation (\ref{HessLem}) is
\begin{gather*}
\nabla_i\nabla_j G +mcg_{ij}G= \partial_{ij}G-\Gamma^k_{ij}\partial_k G+mcg_{ij}G= 0.
\end{gather*}
Dividing by $G$ and setting
$z_k={\partial_k G}/{G}$,
since $ \partial_{jk}G/G=\partial_k z_j+z_kz_j$, we get the f\/irst order PD-system in normal form (Pfaf\/f\/ian system)
with $n$ unknown depending on $n$ variables
\begin{gather*}
\partial_iz_j=-z_iz_j+\Gamma_{ij}^kz_k-mcg_{ij},
\end{gather*}
whose integrability conditions are given by
\begin{gather} \label{int_cond}
\partial_i(-z_jz_l+\Gamma_{jl}^kz_k-mcg_{jl})-\partial_j(-z_iz_l+\Gamma_{il}^kz_k-mcg_{il})=0,
\end{gather}
where the derivatives are computed by considering $z_j$ as functions of $(q^i)$.
These conditions  guarantee that the Pfaf\/f\/ian system is completely integrable, so that there exists a local solution
for any choice of the values of the $(z_j)$ at a point $q_0$, i.e.\ a $n$-dimensional family of solutions parametrized by $(a_1,\ldots, a_n)$ such that
\begin{gather*}
\det\left( \frac{\partial z_j}{\partial a_h} \right) \neq 0.
\end{gather*} Moreover, since $\partial_i z_j=\partial_j z_i$
any solution $z_1,\ldots, z_n$ admits a potential $Z$ such that $z_i=\partial_i Z=\partial_i \ln|G|$. Thus, the solutions $G$ of (\ref{HessLem}) depend also on an additional $(n+1)$-th multiplicative parameter.
We expand (\ref{int_cond}) and by introducing the Riemann tensor
 \begin{gather*}R^k_{lij}=\partial_i \Gamma_{jl}^k -\partial_j \Gamma_{il}^k +\Gamma_{jl}^h\Gamma_{ih}^k
- \Gamma_{il}^h\Gamma_{jh}^k\end{gather*}
and recalling that
\begin{gather*}
\partial_j g_{il}-\partial_i g_{jl}-\Gamma_{jl}^k g_{ik}+\Gamma_{il}^k g_{jk}=0,
\end{gather*}
we get
\begin{gather*}
R^k_{lij}z_k=mc(g_{jl}z_i-g_{il}z_j),
\end{gather*}
that is, since they have to be satisf\/ied for every $z_k$,
\begin{gather*}R^k_{lij}=mc\big(g_{jl}\delta^k_i-g_{il}\delta^k_j\big),\end{gather*}
which, by lowering the index $k$, are the constant curvature conditions
\begin{gather*}
R_{hlij}=K(g_{jl}g_{hi}-g_{il}g_{hj}), \qquad K\in \mathbb{R},
\end{gather*}
and, therefore, $K=mc$.
\end{proof}

\begin{remark}
Since $g^{ij}H_{ij}=\Delta$, where $\Delta$ is the Laplace--Beltrami operator,
the function $G$ satisf\/ies~(\ref{HessTeo}) only if it is an eigenfunction of the Laplace--Beltrami operator of $(Q,\mathbf g)$ with eigen\-va\-lue~$-nmc$, that is a solution of the Helmholtz equation with f\/ixed energy.
The condition is clearly not suf\/f\/icient, because (\ref{HessTeo}) must hold componentwise.
\end{remark}

\begin{proposition} Let $L_i$, $1< i< 2n$, $k-1$ functionally independent first integrals of $L$ in $T^*Q$, let $G$, $U^m(G)$, $V$, $H$ satisfying Theorem~{\rm  \ref{Teo0}}. Then, for any $m\in \mathbb N^*$
\begin{enumerate}\sloppy \itemsep=0pt
\item[$i)$] the $k+2$ functions $(H,U^m(G),L,L_i)$ are functionally independent if and only if $\{L,U^m(G)\}\neq 0$,
\item[$ii)$] if $L$ is a regular natural Hamiltonian on $T^*Q$ and $G(q^i)$ is not constant, then $\{L,U^m(G)\}$ $\neq 0$.
\end{enumerate}
\end{proposition}
\begin{proof}
$i)$ The rank of the Jacobian matrix of the $(H,U^m(G),L,L_i)$ w.r.t.\ the coordinates $(u,p_u,q^i,p_i)$ is equal to the rank of the square $(k+2)\times(k+2)$ matrix
\begin{gather*}
J=
\left(\begin{matrix} \alpha'L & p_u &\alpha \frac{\partial L}{\partial q^a} &\alpha \frac{\partial L}{\partial p_b} \vspace{1mm}\\
\frac {\partial U^m(G)}{\partial u} &\frac {\partial U^m(G)}{\partial p_u} & \frac {\partial U^m(G)}{\partial q^a} & \frac {\partial U^m(G)}{\partial p_b}\vspace{1mm}\\
 0 & 0 & \frac {\partial L}{\partial q^a} & \frac {\partial L}{\partial p_b} \vspace{1mm}\\
  0 & 0 & \frac {\partial L_1}{\partial q^a} & \frac {\partial L_1}{\partial p_b}  \\
\vdots & \vdots & \vdots &   \vdots \\
0 & 0 & \frac {\partial L_{k-1}}{\partial q^a} & \frac {\partial L_{k-1}}{\partial p_b}
\end{matrix}\right)
\end{gather*}
with $|a|+|b|=k$ and where the indices $a$ and $b$ are chosen so that the rank of the $k\times k$ submatrix in the bottom-right corner, which we denote by $J_k$, is $k$. The determinant of $J$ is given by
\begin{gather*}
\det(J)=\left(\alpha'L\frac {\partial U^m(G)}{\partial p_u}-p_u\frac {\partial U^m(G)}{\partial u}\right)\det (J_k).
\end{gather*}
Because $\det (J_k)\neq 0$ by assumption, $\det(J)=0$ if and only if the term between brackets is zero. This term is nothing but $\{H,U^m(G)\}_N$. Therefore, because $\{H,U^m(G)\}=0$, we have $\{H,U^m(G)\}_Q=0$, but this is exactly equivalent to $\{L,U^m(G)\}_Q=\{L,U^m(G)\}=0$.

$ii)$ In the expression $\{L,U^m(G)\}$ the highest-degree term in $p_u$ is
\begin{gather*}
-\sum_{i=1}^n\partial_{p_i}L\partial _{q^i}Gp_u^m.
\end{gather*}
If $L$ is a regular ($\partial_{p_i} L \neq 0$ for all $i$) natural Hamiltonian and $G$ is not constant then the term never vanishes, therefore   $\{L,U^m(G)\}\neq 0$.
\end{proof}

\begin{corollary} If $L$ is a regular natural Hamiltonian on $T^*Q$ and $G$ is not constant, then the functions $(H,U^m(G),L,L_i)$ of the previous proposition are all functionally independent.
\end{corollary}

\begin{corollary} Let $L$ be a Liouville integrable Hamiltonian on $T^*Q$ with $(Q,g)$ a  $($pseudo-$)$Rie\-mannian manifold. If $H$, $L$, $G$ satisfy Theorem~{\rm \ref{Teo0}}, then $H$ is Liouville integrable with additional first integral $U^m(G)$ and, therefore, superintegrable. All additional first integrals of $L$ are also additional first integrals of $H$.
\end{corollary}

In the light of the previous results, in particular Theorem \ref{Teo}, the procedure for constructing  extended Hamiltonians with at least one extra f\/irst integral $U^m(G)$ can be outlined as follows,
\begin{enumerate}\itemsep=0pt
\item Consider a constant-curvature (pseudo-)Riemannian manifold $Q$.

\item  Solve equation (\ref{HessTeo}) for the functions $G(q^i; a_1,\ldots, a_{n+1})$.

\item Solve equation (\ref{VTeo}) for the potential $V$ and build $L$.

\item Determine the extension through Proposition \ref{Ex} and Remark \ref{Ex1} and f\/ix the constant $c$ and the integer $m$ (with the constraint that their product is  the curvature).

\item Compute $U^m(G)$ to obtain the additional f\/irst integrals.
\end{enumerate}

In the following we provide some non trivial examples of the procedure outlined above.

\section{Applications and examples}

\subsection*{Example 1:  $\boldsymbol{n=1}$}

When $Q$ is a one-dimensional Riemannian manifold there are no integrability conditions for the Hessian operator, moreover, after a rescaling of the coordinate, $L$ can be always written as
\begin{gather*}
L=\frac 12 p_v^2+V(v).
\end{gather*}
Hence, the geodesic part of the extended Hamiltonian $H$
\begin{gather*}
\frac{1}{2}\left(p_u^2+\frac{mc}{S_\kappa^2(cu+u_0)}p_v^2\right)
\end{gather*}
corresponds to a Liouville metric on the manifold $N\times Q$, with one Killing vector proportional to $\partial_v$ and constant curvature $K=c^2\kappa$.
The conditions (\ref{HessTeo}) and (\ref{VTeo}) of Theorem \ref{Teo0} become
\begin{gather*}
\frac{d^2}{dv^2}G(v) = mcG(v),\qquad
V'(v)G'(v) = -2mcV(v)G(v).
\end{gather*}
The solution of these dif\/ferential equations are straightforwardly obtained by recalling that
\begin{gather*}
\frac{d}{dx}S_\kappa(x)=C_\kappa(x), \qquad \frac{d}{dx}C_\kappa(x)=-\kappa S_\kappa(x)
\end{gather*}
and they are
\begin{gather*}
G(v)  \propto  S_{mc}(v+v_0),\qquad
V(v)  \propto \frac{1}{C_{mc}^2(v+v_0)},
\end{gather*}
where $\propto$ means ``is a constant multiple of''.
For $c=1$ and $\kappa=0,1,-1$, the manifold $\tilde M=N\times Q$ is respectively,  the Euclidean plane, the sphere $\mathbb S_2$ and the pseudo-sphere $\mathbb H_2$.  For $c=-1$ and $\kappa=0,1,-1$, the Minkowski plane, the deSitter and anti-deSitter  manifolds,  respectively.

\begin{remark} In \cite{CDR3} we obtained for the case $n=1$  a dependence on $m$ for the functions $\alpha$, $G$ and~$V$, while here we have dependence on $m$ for $\alpha$ only. However, a simple rescaling, that for this case is given by $v\rightarrow  \frac vm$, shows that the two results are equivalent. This suggests that the most natural way to write the Tramblay--Turbiner--Winternitz like systems considered in \cite{CDR3,BKM1,BKM2,MPY,TTW1,TTW2} is probably to shift the dependence on $m$ from the potential function~$V$ of~$L$ to the metric factor $\alpha$.  A similar rescaling occurs for higher dimensional~$L$. If $L$ includes trigonometric functions of the $(q^i)$, the rescaling makes evident the dihedral, polyhedral  or, in general, discrete symmetries of~$L$.
\end{remark}

\subsection*{Example 2:  $\boldsymbol{n=2}$}

{\bf The f\/lat case. }
If $(Q,\mathbf g)$ is a f\/lat manifold with Cartesian coordinates $(q^1,q^2)$, a nonvanishing solution for $G$ can be obtained by assuming $c=0$ (see Lemma~\ref{PropClaudia}) and therefore
$\gamma=-A(u+u_0)$ (Lemma~\ref{Ex}). In this case, equation \eqref{HessTeo} becomes $\partial_1\partial_2 G=0$ and its solution is $G=k_0+k_1q^1+k_2q^2$, with $k_i$ not simultaneously zero. The compatible potential $V$ obtained from equation \eqref{VTeo} is{\samepage
\begin{gather*}
V= mL_0\left[ \left( q^1+\frac {k_0}{2k_1}\right)^2+\left( q^2+\frac {k_0}{2k_2}\right)^2\right]+F\big(k_1q^2-k_2q^1\big),\qquad  \text{if} \quad k_1k_2 \neq 0,\\
V= mL_0\left(q^1+\frac {k_0}{k_1}\right)^2+F\big(q^2\big),\qquad  \text{if} \quad k_2=0.
\end{gather*}
The two forms of $V$ are equivalent up to rotations.}

{\bf The pseudo-sphere.}
If $(Q,\mathbf g)=\mathbb H_2$, the 2-dimensional pseudosphere of curvature $K=-1$, with $g^{11}=1$, $g^{22}=4 (e^\eta+e^{-\eta})^{-2}=\cosh^{-2}(\eta)$ in orthogonal coordinates $(\eta, \xi)$, we have $mc=-1$ and from (\ref{HessTeo})
\begin{gather*}
G=\big(a_1+a_2e^\xi+a_3e^{-\xi}\big)e^{-\eta}+\big(a_2e^\xi+a_3e^{-\xi}-a_1\big)e^\eta.
\end{gather*}
For $a_2=a_3=0$ the integration of (\ref{VTeo}) gives for the potential
\begin{gather*}
V=  \frac{F(\xi)}{\cosh^2(\eta)}.
\end{gather*}
In this case $V$ is  in St\"ackel form $V=g^{ii}f_i(q^i)$ and  compatible with the separation of variables
of $L$, which is therefore integrable with one quadratic f\/irst integral $H_1$ at least. The extended Hamiltonian $H$ is therefore always superintegrable with the four f\/irst integrals $H$, $L$, $H_1$ and $U^m(G)$ for $mc=-1$. We write explicitly $U^m(G)$, $m=1,2$, relatively to the extension
\begin{gather*}
H=\frac 12 p_u^2-\frac {m^2}{u^2}L, \qquad \gamma=-\frac mu,
\end{gather*}
as follows
\begin{gather*}
U(G)=-2\sinh(\eta)p_u+2\frac{\cosh (\eta)}u p_1,
\\
U^2(G)=-2\sinh(\eta)p_u^2+8\frac{\cosh(\eta)}up_up_1-16\frac{\sinh(\eta)}{u^2} L.
\end{gather*}

{\bf The sphere.}
If $(Q,\mathbf g)=\mathbb S^2$, the 2-dimensional sphere of curvature $K=1$, with $g^{11}=1$, $g^{22}=\sin^{-2}\theta)$ in standard spherical coordinates $(\theta, \phi)$, we have $mc=1$ and from (\ref{HessTeo})
\begin{gather*}
G= (a_1\sin\phi+a_2 \cos\phi)\sin\theta+a_3\cos\theta.
\end{gather*}
For $a_3=0$,  the integration of (\ref{VTeo}) gives for the potential
\begin{gather*}
V=\frac 1{\cos ^2\theta}F\left((a_1\cos \phi-a_2\sin \phi)\tan \theta \right),
\end{gather*}
which in general is not in St\"ackel form and therefore we have an example of a non separable~$L$ which admits
an extension (again not separable) with a polynomial additional f\/irst integral. For $m=1$ and $F=(a_1\cos \phi-a_2\sin \phi)\tan \theta$, for example, we have with $V$ of above
\begin{gather*}
U(G)=(a_1\sin \phi+a_2\cos \phi)\left(p_u\sin \theta+\frac 1u p_\theta \cos \theta\right)+\frac 1{u\sin \theta}p_\phi(a_1\cos \phi-a_2\sin \phi).
\end{gather*}

\subsection*{Example 3:  $\boldsymbol{n=3}$}
Let us consider as $(Q,\mathbf g)$ the sphere $\mathbb S^3$ with coordinates ($q^1=\eta$, $q^2=\xi_1$, $q^3=\xi_2$) where $0<\eta<\pi/2$ and $0\leq \xi_i <2\pi$
and the parameterization in $\mathbb R^4$ being given by
\begin{gather*}
x = \cos\xi_1 \sin\theta,\qquad
y =  \sin\xi_1 \sin\theta,\qquad
z =  \cos\xi_2 \cos\theta, \qquad
t = \cos\xi_2 \cos\theta.
\end{gather*}
These coordinates are known as Hopf coordinates, the non null components of the covariant metric tensor are $g_{11}=1$, $g_{22}=\sin^2\eta$ and
$g_{33}=\cos^2\eta$ so that the curvature is $K=1$. The surfaces $\eta={\rm const}$ are f\/lat tori spanned by the rotations $\partial_{\xi_i}$ which are Killing vectors of the manifold. These coordinates correspond to a cylindrical rotational separable system, they are associated with a Killing 2-tensor $\mathbf K$  which is described in appendix in~\cite{CMS} and which provides a quadratic f\/irst integral of the geodesics $H_1=\frac 12 K^{ij}p_ip_j$. Therefore, the geodesic Hamiltonian $G=\frac 12 g^{ii}p_i^2$ of $\mathbb S_3$ admits the following four independent quadratic in the momenta f\/irst integrals $G$, $H_1$, $p_2^2$, $p_3^2$ and is Liouville integrable. By applying our procedure we obtain from~(\ref{HessTeo})
\begin{gather*}
G=(a_3\sin \xi_1 +a_4\cos \xi_1 )\sin \eta +(a_1\sin \xi_2 +a_2\cos \xi_2 )\cos \eta ,
\end{gather*}
with $a_i$ constants. After setting $a_2=a_3=a_4=0$ one obtains easily from (\ref{VTeo})
\begin{gather*}
V= \frac{1}{\sin^2\eta}F\left(\xi_1,\frac{\tan\eta}{\cos \xi_2}\right).
\end{gather*}
Remarkably, this potential can be compatible with separation of variables, then making $G+V$ an integrable system, if it is  in St\"ackel form $V=g^{ii}f_i(q^i)$ as happens for example with $F=F(\xi_1)$. However, in general it is not, as for
\begin{gather}\label{Fno}
F=\frac{\sin \xi_1 }{\cos \xi_2}\tan\eta.
\end{gather}
In both cases our procedure provides independent f\/irst integrals $U^m(G)$ for the extended Hamiltonians $H$ described in the previous sections, with the prescription $mc=1$. In the case of $F=F(\xi_1)$, for example, $H$ is a superintegrable Hamiltonian with, at least,  four other independent f\/irst integrals. In the case (\ref{Fno}), $H$ admits the two other independent f\/irst integrals~$L$ and~$U^m(G)$ at least. For $V$ with $F$ given by(\ref{Fno}), we have for example
\begin{gather*}
U(G)=p_u\cos \eta \sin \xi_2+\frac 1u\left( p_3\frac{\cos \xi_2}{\cos \eta} -p_1\sin \eta \sin\xi_2 \right).
\end{gather*}

\section{Conclusions}
We have proved that the method developed in \cite{CDR3} for building polynomial additional f\/irst integrals of arbitrarily high order of extended two-dimensional Hamiltonians   can be generalized to similar extensions of $n$-dimensional Riemannian or pseudo-Riemannian manifolds. As a consequence of natural integrability conditions,  it is  necessary  that  the  curvature  of these manifolds is constant. In  examples for $n=1,2,3$ we improve the results obtained earlier for the one-dimensional manifolds and show how the procedure works in the two and three dimensional ones. Future directions of research will be towards the characterisation of more general procedures of extension of Hamiltonian systems and the search for more general expressions of the operator $U$ and of the function~$G$.

\subsection*{Acknowledgements}

The research has been partially supported (C.C.) by the European program  ``Dote ricercatori'' (F.S.E. and Regione Lombardia).
G.R. is particularly grateful to the University of Waterloo, ON, Canada, where part of the research has been done during a visit.
The authors wish to thank F.~Magri and R.G.~McLenaghan for their suggestions and stimulating discussions about the topic of the present research.

\pdfbookmark[1]{References}{ref}
\LastPageEnding

\end{document}